# Data-Driven Scheduling of Electric Boiler with Thermal Storage for Providing Power Balancing Service

Likai Liu
Department of Engineering Engineering
Tsinghua University
Beijing, China
llk17@mails.tsinghua.edu.cn

Zechun Hu
Department of Engineering Engineering
Tsinghua University
Beijing, China
zechhu@tsinghua.edu.cn

Jian Ning
North China Branch of State Grid Corporation of China
State Grid Corporation of China
Beijing, China
ning.jian@nc.sgcc.com.cn

Yilin Wen
Department of Engineering Engineering
Tsinghua University
Beijing, China
wen-yl20@mails.tsinghua.edu.cn

*Abstract*—**The rapid development of renewable energy has increased the peak to valley difference of the netload, making the netload follwing being a new challenge to the power system. Electric boiler with thermal storage (EBTS) occupies a non-negligible part of the load in the winter season in Northern China. EBTS operation optimization can not only save its own energy cost but also reduce the peak shaving and valley filling pressure of the system. To this end, the operation optimization of EBTS for providing the power balancing service is studied in this paper, which mainly includes three parts: First, the joint probability distribution between the predicted and actual temperatures is built by utilizing the Copula theory; Secondly, the actual temperatures are sampled based on the predicted temperatures of the next day, and the scenario set is generated by clustering these samples, where *K*-means clustering method are used; Thirdly, the stochastic operation optimization model of EBTS considering the uncertainty of outdoor temperature is constructed. Through the case study, it is found that the proposed method can save the total operation cost of the EBTS compared with the deterministic EBTS operation optimization model.**

## I. Introduction

The high penetration of renewable energy has largely increased the peak to valley difference of the power system netload. Moreover, the load following reserve has also decreased because fewer traditional generators are scheduled. As a result, the peak shaving and valley filling pressure has soared [1], and some power systems cannot absorb all the renewable energies for lack of load following reserve. Northern China faces greater pressure for power balancing at night in winter: the wind speed is relatively high at this time, but a large number of thermal power plants cannot lower their power output too much because they need to supply heat to the resident.

To alleviate the challenge of power balancing and promote the consumption of renewable energy, lots of researchers has studied utilizing flexible loads to provide power balancing

This work is supported by the science and technology project of state grid corporation of China.(2600/2020-02001B)

service [2]. These researches can be divided into four kinds according to the types of flexible resources, namely energy storage devices, electrical vehicles, controllable thermal loads, and virtual power plants. Electric boiler with thermal storage (EBTS) is a typical controllable thermal load, and it takes a remarkable part of the load in Northern China in winter season. The utilization of thermal storage boiler as a flexible resource can effectively relieve the netload following pressure in Northern China.

Many pieces of research about the coordinated operation of combined heat and power units, thermal storage boiler, and wind power has been published recently [3]–[5]. The centralized dispatch of the integrated energy system can effectively increase the power system flexibily, which is beneficial to the overall social interest. Nevertheless, each participator of the integrated energy system belongs to different interest groups. Therefore, the collaborative optimization between them may face challenges in practice. In reference [6], a master-slave game trading method is proposed to jointly operate the wind power plant and EBTS, which is faced to the bilateral trade of the power balancing auxiliary service.

The power balancing auxiliary service market has developed rapidly recently in China, but the studies on the operation optimization of EBTS for providing power balancing service are few. In this study, a stochastic operation optimization model of EBTS for providing the power balancing service is proposed. The uncertainty of outdoor temperature is considered by utilizing Copula theory to build the joint distribution of predicted and actual outdoor temperatures, and the scenario set of the outdoor temperatures are generated by clustering the samples taken from the joint distribution model.

The remainder of the paper is organized as follows. The method for modeling the uncertainty of outdoor temperatures is developed in Section II. Section III builds the formulation for the EBTS operation optimization model. Numerical experiments are presented in Section IV. Section V concludes this paper.

## II. UNCERTAINTY MODELING OF OUTDOOR TEMPERATURE

The outdoor temperature clearly influences the thermal load of the heating station. Nevertheless, the temperature forecast only has the pointwise prediction, which only provides very limited information for the EBTS operation. By using the historical data of the predicted and actual temperatures, the joint distribution of them can be established. Then, the conditional distribution of the actual temperature under the predicted one can be obtained. The probability distribution will provide more comprehensive information about the future temperature for the EBTS operation optimization.

In this research, the Copula theory and $K$-means clustering method are utilized cooperatively to generate the scenario set for the stochastic EBTS operation optimization model, as shown in Fig. 1, which are introduced in this section.

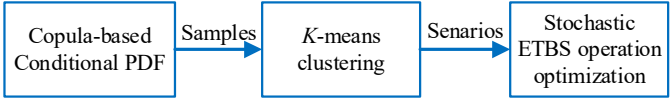

Fig. 1. Flow chart of the proposed method.

### A. Copula-Based Joint Distribution Establishment

Copula theory provides an effective way to construct a multivariate distribution [7], which transforms the modeling of the joint cumulative distribution into the modeling of marginal cumulative distributions and the Copula function separately. By utilizing the Copula function, the joint cumulative distribution function (CDF) of the forecasted and actual outdoor temperatures denoted by $\overline{T^{\text{env}}}$ and $T^{\text{env}}$ can be expressed as follows:

$$\mathcal{F}_{\overline{T^{\text{env}}}, T^{\text{env}}}\left(\overline{T^{\text{env}}}, T^{\text{env}}\right) \\ = \mathcal{C}_{\overline{T^{\text{env}}}, T^{\text{env}}}\left(\mathcal{F}_{\overline{T^{\text{env}}}}\left(\overline{T^{\text{env}}}\right), \mathcal{F}_{T^{\text{env}}}\left(T^{\text{env}}\right)\right), \quad (1)$$

For clarity, a Copula CDF and corresponding probability distribution function (PDF) are given in Appendix A as an example.

There are three steps in constructing the joint distribution model by utilizing the Copula theory: first, build the marginal CDF of each stochastic variable; secondly, calculate the rank correlation coefficient between the variables by using the historical data; thirdly, calculate the parameters in the Copula function according to the rank correlation coefficient. The detailed process is omitted due to the limited space, the interested readers are referred to [7] for the detail.

The joint PDF $f_{\overline{T^{\text{env}}}, T^{\text{env}}}$ is the derivative of the joint CDF $\mathcal{F}_{\overline{T^{\text{env}}}, T^{\text{env}}}$ on the forecasted and actual temperatures, given as follows:

$$f_{\overline{T^{\text{env}}}, T^{\text{env}}}\left(\overline{T^{\text{env}}}, T^{\text{env}}\right) = \frac{\partial \mathcal{F}_{\overline{T^{\text{env}}}, T^{\text{env}}}\left(\overline{T^{\text{env}}}, T^{\text{env}}\right)}{\partial \overline{T^{\text{env}}} \partial T^{\text{env}}} \\ = \varsigma_{\overline{T^{\text{env}}}, T^{\text{env}}}\left(\mathcal{F}_{\overline{T^{\text{env}}}}\left(\overline{T^{\text{env}}}\right), \mathcal{F}_{T^{\text{env}}}\left(T^{\text{env}}\right)\right) \\ \cdot f_{T^{\text{env}}}\left(T^{\text{env}}\right) \cdot f_{\overline{T^{\text{env}}}}\left(\overline{T^{\text{env}}}\right) \quad (2)$$

where $\varsigma_{\overline{T^{\text{env}}}, T^{\text{env}}}$ is the Copula density function for the joint PDF $f_{\overline{T^{\text{env}}}, T^{\text{env}}}$.

Then, the conditional PDF of the actual temperature under the predicted one can be obtained by dividing the PDF of the predicted temperature into the joint PDF of forecasted and actual temperatures, as:

$$f_{T^{\text{env}}|\overline{T^{\text{env}}}}\left(T^{\text{env}}|\overline{T^{\text{env}}}\right) \\ = \varsigma_{\overline{T^{\text{env}}}, T^{\text{env}}}\left(\mathcal{F}_{\overline{T^{\text{env}}}}\left(\overline{T^{\text{env}}}\right), \mathcal{F}_{T^{\text{env}}}\left(T^{\text{env}}\right)\right) \cdot f_{T^{\text{env}}}\left(T^{\text{env}}\right) \quad (3)$$

### B. Scenario Generation of the Outdoor Temperature

After getting the conditional PDF (3), samples of the actual temperature can be taken based on the predicted temperature. Then, the scenario of the outdoor temperature can be generated by utilizing the clustering method.

In this research, the $K$-means clustering method is utilized, which is one of the most widely used clustering methods. $K$-means clustering is a type of unsupervised learning, which iteratively executes two steps and finally assigns each data sample to one of the $K$ groups. At the beginning of this algorithm, each group centroid are randomly generated. The first step is to assign each data point to its nearest centroid, and the distance adopts the Euclidean distance, as:

$$\underset{c_i \in C}{\arg\min} \operatorname{dist}(c_i, x)^2 \quad (4)$$

where $c_i$ is the center of the group $i$, $C$ is the set of clustering centroids, and $x$ is the sample.

The second step is to update the centroids by utilizing the result of data assigment, which calculates the centriod of each group of data samples, as:

$$c_i = \frac{1}{|S_i|} \sum_{x_i \in S_i} x_i \quad (5)$$

where $S_i$ is the set of all samples belonging to group $i$.

In the $K$-means clustering, the number of the group should be stipulated in advance. Nevertheless, there is no method for determining the exact value of $K$. The general method is to run the $K$-means clustering algorithm for a range of $K$ values and compare the results. One of the commonly used metrics is the mean distance between data points to their cluster center. Since this metric will not increase with the increase of $K$, the change rate is used to determine the value of $K$. The criterion is that choosing a $K$ where the change rate sharply shifts in $K$+1.

After clustering, $K$ scenarios can be obtained and will be used to build the stochastic operation optimization model of EBTS for providing power balancing service.

## III. EBTS OPERATION OPTIMIZATION MODEL

### A. Objective Function

The objective function of the EBTS operation optimization model contains the electricity cost, the revenue from providing power balancing reserve, and default penalty. The electricity cost is the product of time-of-use electricity price and power consumption. The revenue is calculated based on compensation unit price and regulation capacity, which is the difference between the day-ahead bade power consumption and the power baseline recognized by the grid. The default penalty is determined based on the actual power consumption and the day-ahead bade power consumption, and the deviation within the interval $[-\epsilon, \epsilon]$ is free of penalty. In summary, the objective function for the EBTS operation optimization model is formulated as follows:

$$F = \min \sum_{n=1}^{N} \left\{ C_n^{\mathrm{E}} \cdot P_n - C_n^{\mathrm{R}} \cdot (O_n - B_n) + C_n^{\mathrm{F}} \cdot \left[ (P_n - O_n - \epsilon)^+ + (O_n - P_n - \epsilon)^+ \right] \right\} \quad (6)$$

where $n$ and $N$ are the index and total number of the hours in one day, $P_n$ is the actual power consumption in hour $n$, $C_n^E$ is the time-of-use electricity price of hour $n$, $C_n^R$ is the compensation unit price of hour $n$, $B_n$ is the power baseline in hour $n$, $O_n$ is the day-ahead bade power consumption in hour $n$, $C_n^{\mathrm{F}}$ is the unit price for the penalty of power deviation in hour $n$, the function $[x]^+$ is max($x$,0).

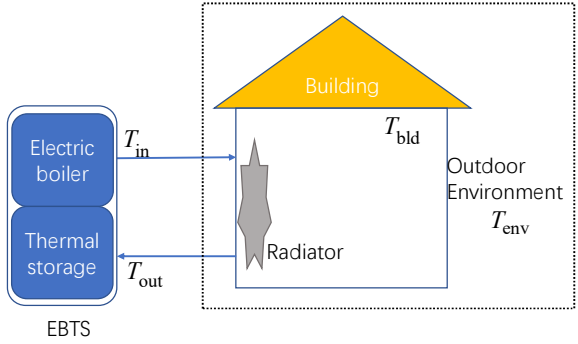

Fig. 2. Schematic of heating device and heat load.

As shown in the Fig. 2, after a certain simplification, the system studied in this paper consists of four parts: the thermal storage tank, electric boiler, several buildings, and the external environment. Each building is assumed to has one radiator in this study.

### B. Operation Constraints of EBTS

*1) Power Balance Constraint of EBTS:* An EBTS is typically composed of electric boilers and thermal storage tanks. In order to increase the versatility of the model, this study assumes that the EBTS operates in a combination way of direct heat supply and thermal storage supply, which specifically includes four working modes: 1) The boiler directly supplies heat to the building; 2) The boiler supplies some heat to the building and store the other heat to the thermal storage tank; 3) The thermal storage tank solely supplies heat to the building; 4) The boiler and thermal storage tank simultaneously supply heat to the building. The above processes can be expressed by the following formula:

$$P_n \cdot \eta - P_n^{\mathrm{str}} + P_n^{\mathrm{rls}} = \sum_{m \in \mathcal{M}} D_{m,n}, \forall 1 \leq n \leq N \quad (7)$$

where $\eta$ is the conversion efficiency from electrical energy to heat energy, $P_n^{\mathrm{str}}$ is the equivalent power of the heat energy stored into the thermal storage during hour $n$, $P_n^{\mathrm{rls}}$ is the equivalent power released by the thermal storage during hour $n$, $D_{m,n}$ is the equivalent power of heat spplied to building $m$ during hour $n$, $\mathcal{M}$ is the set of buildings requiring heat supply.

*2) Operational Range Constraints of EBTS:* All the power values $P_n$, $P_n^{\mathrm{str}}$, and $P_n^{\mathrm{rls}}$ should be within their operational ranges, as:

$$P_{\min} \leq P_n \leq P_{\max}, \forall 1 \leq n \leq N. \quad (8)$$

$$P_{\min}^{\mathrm{str}} \leq P_n^{\mathrm{str}} \leq P_{\max}^{\mathrm{str}}, \forall 1 \leq n \leq N. \quad (9)$$

$$P_{\min}^{\mathrm{rls}} \leq P_n^{\mathrm{rls}} \leq P_{\max}^{\mathrm{rls}}, \forall 1 \leq n \leq N. \quad (10)$$

where $P_{\max}$ and $P_{\min}$ are the upper and lower limits of the power consumption of EBTS, $P_{\max}^{\mathrm{str}}$ and $P_{\min}^{\mathrm{str}}$ are the upper and lower limits of power stored into the thermal storage, and $P_{\max}^{\mathrm{rls}}$ and $P_{\min}^{\mathrm{rls}}$ are the upper and lower limits of power released by the thermal storage.

*3) Energy Balance Constraint of Thermal Storage:* The thermal storage tank is an important part of the EBTS. In actual operation, in order to ensure the thermal storage tank keeping the maximum heat storage capacity, the water volume in the tank will be maintained at the maximum level [8] all the time. Therefore, the thermal energy stored in the thermal storage tank is mainly related to the water temperature. The energy balancing constraint considering the thermal energy dissipation can be formulated as:

$$H^n = H^{n-1} \cdot v + P_n^{\mathrm{str}} \cdot \Delta t - P_n^{\mathrm{rls}} \cdot \Delta t, \quad (11)$$

where $H^n$ is the thermal energy stored in the water tank at the beginning of hour $n$.

*4) Operational Range Constraints of Thermal Storage:* The thermal energy stored in the water tank should be kept within the operational range:

$$H_{\min} \leq H^n \leq H_{\max}, \forall 1 \leq n \leq N \quad (12)$$

where $H_{\max}$ and $H_{\min}$ are the upper and lower limits of the thermal energy stored in the water tank.

### C. Operation Constraints of Radiator

*1) Power Balance Constraint of Radiator:* In this research, it is assumed that the heating system is small, and outlet water temperature of the heating station is equal to the inlet water temperature of the building. Then, the equivalent power of the radiator is equal to the thermal energy supplied by the EBTS, which can be expressed by the mass flow rate of hot water $\dot{M}_{\mathrm{hot}}$, the specific heat capacity of the water $c_{\mathrm{hot}}$, and the temperature of the inlet and outlet water of the building ($T^{\mathrm{in}}$ and $T^{\mathrm{out}}$) as follows:

$$D_{m,n} = \dot{M}_m c_{\mathrm{hot}} \left( T_{m,n}^{\mathrm{in}} - T_{m,n}^{\mathrm{out}} \right) \quad (13)$$

There exists heat exchange between the radiator and the building, and a relationship among the temperature of the inlet and outlet water of the building, as well as the indoor temperature of the building can be obtained as follows:

$$T_{m,n}^{\text{out}} = (1 - \vartheta_m)T_{m,n}^{\text{in}} + \vartheta_m T_{m,n}^{\text{bld}} \quad (14)$$

$\vartheta_m$ is the coefficient reflecting the overall heat dissipation effect of the radiator, which is calculated as follows:

$$\vartheta_m = 1 - \exp\left(\frac{-\mathcal{K}F}{\dot{M}_m^{\text{hot}} c_{\text{hot}}}\right) \quad (15)$$

where $\mathcal{K}$ is the heat dissipation coefficient of the radiator, $F$ is the heat dissipation area of the radiator.

*2) Temperature Constraint of Radiator:* The radiator has temperature resistance limitations, so the temperatures of its inlet and outlet water cannot be too high or too low:

$$T_{\min}^{\text{in}} \leq T_{m,n}^{\text{in}} \leq T_{\max}^{\text{in}}, \forall 1 \leq n \leq N \quad (16)$$

$$T_{\min}^{\text{out}} \leq T_{m,n}^{\text{out}} \leq T_{\max}^{\text{out}}, \forall 1 \leq n \leq N \quad (17)$$

where $T_{\max}^{\text{in}}$ and $T_{\min}^{\text{in}}$ are the upper and lower limits of inlet water temperature, $T_{\max}^{\text{out}}$ and $T_{\min}^{\text{out}}$ are the upper and lower limits of outlet water temperature.

### D. Constraints Related to the Building

*1) Energy Balance Constraint of Building:* The building receives energy from the radiator and exchanges heat with the outdoor environment. Its temperature change can be expressed as the following formula:

$$c_{\text{bld}}^m (T_{m,n+1}^{\text{bld}} - T_{m,n}^{\text{bld}}) = U_{\text{bld}} \left(T_n^{\text{env}} - T_{m,n}^{\text{bld}}\right) + D_{m,n} \quad (18)$$

where $c_{\text{bld}}^m$ is the equivalent specific heat capacity of the building $m$, $T_{m,n}^{\text{bld}}$ is the initial temperature of the building $m$ in the $n^{th}$ hour, $U_m$ is the heat conductivity coefficient between the building $m$ and the outdoor environment, and $T_n^{\text{env}}$ is the initial temperature of the environment in the $n^{th}$ period.

*2) Temperature Constraint of Building:* Considering the resident's amenity on the indoor temperature, the building temperature should be kept within an interval:

$$T_{\min}^{\text{bld}} \leq T_{m,n}^{\text{bld}} \leq T_{\max}^{\text{bld}}, \forall 1 \leq n \leq N \quad (19)$$

where $T_{\max}^{\text{bld}}$ and $T_{\min}^{\text{bld}}$ are the upper and lower limits of indoor temperature.

### E. Solution Method

Based on the simplified steady state model of the heating supply system, the objective function and all of the constraints in the estainblished optimization model are linear, and there are only continuous variables in this model. Thus, this model is a linear programming problem, which is solved by Gurobi [9] in this work. Besides, the case study is implemented in MATLAB and programmed using YALMIP [10].

## IV. CASE STUDY

### A. Case Settings

In this research, the heating station has three electrode-type electric boilers and three thermal storage tanks. The rated power of each electric boiler is 20 MW, and the conversion efficiency from electricity to heat is 99%. The maximum heat storage capacity for each thermal storage tank is 100.25 MWh, the maximum heat storage power for a single thermal storage tank is 75 MW, and the heat loss is 5% per-hour. The time-of-use power price is given in Table I.

TABLE I
TIME-OF-USE POWER PRICE

| Load | Time | Power price |
| --- | --- | --- |
| Peak | 11:00-12:00, 19:00-21:00 | 0.9792 |
| High | 9:00-11:00, 18:00-19:00, 21:00-23:00 | 0.8603 |
| Normal | 7:00-9:00, 12:00-18:00 | 0.6223 |
| Trough | 23:00-7:00 | 0.3843 |

The thermal parameters of the buildings are given in Table II, which are taken from subsection 6.4.3 in reference [11]. The inlet water temperature range is 60°C-90°C, the outlet water temperature range is 20°C-40°C, and the indoor temperature range is 18°C-24°C.

TABLE II
THERMAL PARAMETERS ABOUT THE BUILDING

| Building index | Specific heat capacity (10GJ/°C) | Thermal conductance (10kW/°C) | Mass flow rate (kg/s) | Parameter $\vartheta$ |
| --- | --- | --- | --- | --- |
| 1 | 4.256 | 1.079 | 11.832 | 0.809 |
| 2 | 4.394 | 1.485 | 16.221 | 0.801 |
| 3 | 3.216 | 1.479 | 18.185 | 0.754 |
| 4 | 4.405 | 1.243 | 14.335 | 0.804 |
| 5 | 3.980 | 1.400 | 15.812 | 0.764 |
| 6 | 3.171 | 1.071 | 12.355 | 0.769 |
| 7 | 3.445 | 1.211 | 13.703 | 0.784 |
| 8 | 3.851 | 1.458 | 17.256 | 0.772 |
| 9 | 4.472 | 1.396 | 16.072 | 0.794 |
| 10 | 4.483 | 1.480 | 17.68 | 0.776 |

The historical temperature data are taken from Zhangjiakou, a prefecture-level city in northwestern Hebei province in Northern China. The data from October 1, 2016 to July 1, 2020 is recorded every 3 hours, and the data from July 2, 2016 to April 20, 2021 is recorded every hour. The heat supply period for Zhangjiakou is from November 1st to April 1st of the next year. This study uses historical temperature prediction data and measured data from the winter heating period of 2016 to the spring heating period of 2020 to construct the Copula-based joint probability distribution model, and uses the data of the heating period from the winter of 2020 to the spring of 2021 to test the proposed method. We selected two days in every months in the heating period for test in this paper.

### B. Copula Model Selection

There are many kinds of Copula function in Copula theory, and five Copula function, namely Gaussian Copula, Student-t

TABLE III
PERFORMANCES OF DIFFERENT COPULA MODELS

| Model | Gaussian | Student-t | Gumbel | Clayton | Frank |
|---|---|---|---|---|---|
| BIC ($10^4$) | -2.0024 | -2.0025 | -1.9294 | -1.8461 | -1.5953 |

TABLE IV
COMPUTATION TIME FOR DIFFERENT METHODS

| Method | Proposed Method | Deterministic optimization |
|---|---|---|
| Coputation time | 0.39 s | 0.11 s |

Copula, Clayton Copula, Gumbel Copula, and Frank Copula are used as the candidate Copula function to construct the joint distribution model in this study. The model performances are compared by utilizing the Bayesian information criterion (BIC) [12], whose definition is:

$$BIC = -2\ln(\delta) + q * \ln(\mathcal{X}) \quad (20)$$

where $q$ denotes the number of undetermined parameters in a Copula model, $\delta$ indicates the value of the maximum likelihood function [13], and $\mathcal{X}$ is the total number of samples used to build the joint PDF. The performances of each model are shown in Table III. The Student-t Copula model is chosen to build the joint distribution model because it has the samllest BIC.

### D. Method Comparison

To show the superiority of the proposed method, the proposed stochastic optimization model for EBTS operation is compared with the deterministic optimization model utilizing the pointwise temperature prediction as the input, which equate with the stochastic optimization model with only one scenario. These two methods are compared from the optimization result and computational efficiency.

*1) Comparison on the optimization results:* The optimization results of different methods are shown in Fig. 5, where the total costs of the proposed method are no larger than those of the deterministic optimization model for every test day. The mean operation costs for the stochastic and deterministic optimization model are $1.40 * 10^5$ and $1.42 * 10^5$ China Yuan respectively. The comparison results illustrate that the proposed method is superior than the deterministic method.

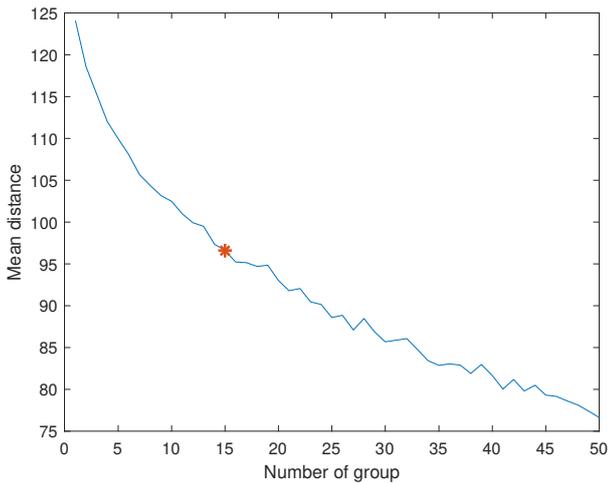

Fig. 3. Mean distances from samples to their clustering centers under different $K$.

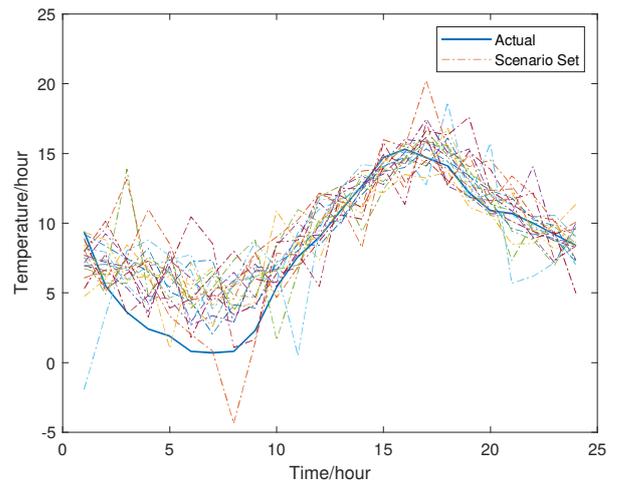

Fig. 4. Actual temperature and its scenario set used in the stochastic optimization.

### C. K-Means Clustering

In order to choose the suitable $K$ for $K$-means clustering method, the clustering results under different $K$ are compared. First, one day's predicted temperatures are selected, and 400 samples of the actual temperatures are taken from (3). Then, these samples are clustered by using different $K$ (from 1 to 50 with a step-size of 1). The mean distances of the data sample to the clustering center under different values of $K$ are plotted in Fig. 3. It can be found that the mean distance decrease sharply before $K = 15$, then the decrease rate slows down gradually. Thus, $K$ takes 15 in this research. The clustering result for one selected day is shown in Fig. 4.

*2) Comparison on the computational efficiency:* The experiments are performed on a PC with an Intel(R) Core(TM) i7-7700 CPU 3.6 GHz and 8 GB of memory. As shown in Table IV, the computation time of the proposed stochastic optimization model is longer than the deterministic model. However, the operation optimization of EBTS is executed every 24 hours; thus, the computation time of the proposed method is fully acceptable.

## V. CONCLUSION

This paper proposed a data-driven stochastic optimization model of EBTS for providing the power balancing service. The Copula theory is utilized to build the joint distribution of predicted and actual outdoor temperatures. Then, the scenario set of the actual outdoor temperatures are generated by clustering the samples taken from the Copula-based joint distribution model. The simulation results show that the proposed method can save the total operational cost compared with the deterministic optimization method.

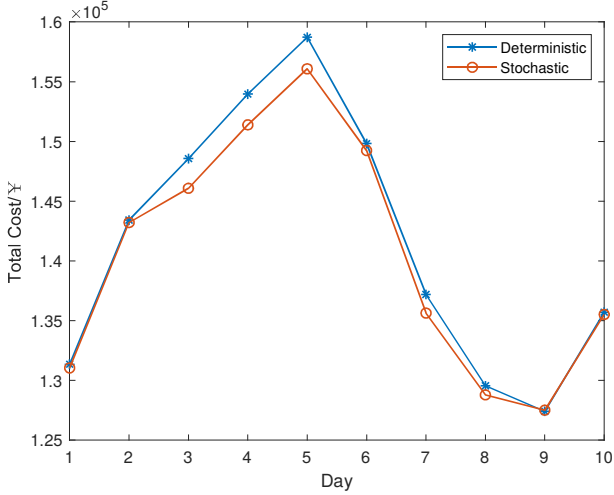

Fig. 5. Optimization results of the deterministic and stochastic optimization model.

The scenario-based stochastic optimization method may get the poor out-of-sample performance when the estimation on the distribution of the stochastic variables has comparatively large error. Future research focuses on improving the robustness of the optimization model to the error on the estimated distribution of the outdoor temperature.

## APPENDIX A

Multivariate Gaussian Copula is a widely used Copula function. Let $\varpi$ be a symmetric, positive definite matrix with diag($\varpi$) =1. $\Psi_\varpi$ be the standardized multivariate normal distribution with correlation matrix $\varpi$, and $\Psi^{-1}$ be the inverse function of the normal distribution. The multivariate Gaussian Copula for the stochastic variables $\{u_1, ..., u_z, ..., u_Z\}$ that follow the uniform distribution is then defined as follows [7]:

$$\begin{aligned} &\mathcal{C}\left(u_1, ..., u_z, ..., u_Z; \varpi\right) = \\ &\Psi_\varpi\left(\Psi^{-1}\left(u_1\right), ..., \Psi^{-1}\left(u_z\right), ..., \Psi^{-1}\left(u_Z\right)\right) \end{aligned} \quad (21)$$

And the corresponding density function is:

$$\begin{aligned} &\varsigma\left(u_1, \ldots, u_z, \ldots, u_Z; \varpi\right) = \\ &\frac{1}{|\varpi|^{1/2}} \exp\left(-\frac{1}{2}\psi^\top \left(\varpi^{-1} - I_Z\right)\psi\right) \end{aligned} \quad (22)$$

where $\psi$ is a vector composed of $\psi_z$, with $\psi_z = \Psi^{-1}(u_z)$, and $I_Z$ is the unit matrix with dimension of $Z$.